\documentclass[twocolumn,english,prb]{revtex4}
\usepackage[T1]{fontenc}
\usepackage[latin9]{inputenc}
\usepackage[a4paper]{geometry}
\usepackage{amstext}
\usepackage{graphicx}
\usepackage{amssymb}

\makeatletter

\providecommand{\tabularnewline}{\\}

\@ifundefined{textcolor}{}
{%
 \definecolor{BLACK}{gray}{0}
 \definecolor{WHITE}{gray}{1}
 \definecolor{RED}{rgb}{1,0,0}
 \definecolor{GREEN}{rgb}{0,1,0}
 \definecolor{BLUE}{rgb}{0,0,1}
 \definecolor{CYAN}{cmyk}{1,0,0,0}
 \definecolor{MAGENTA}{cmyk}{0,1,0,0}
 \definecolor{YELLOW}{cmyk}{0,0,1,0}
 }

\makeatother

\usepackage{babel}

\begin{document}

\title{Spin coupling around a carbon atom vacancy in graphene}

\author{M. Casartelli$^{1}$, S. Casolo$^{1}$, G. F. Tantardini$^{1,2}$,
R. Martinazzo$^{1,2}$}

\affiliation{$^{1}$Dipartimento di Chimica, Università degli Studi di Milano,
via Golgi 19, 20133 Milan, Italy}

\affiliation{$^{2}$Istituto di Scienze e Tecnologie Molecolari, CNR, via Golgi
19, 20133 Milan, Italy}

\email{rocco.martinazzo@unimi.it}
\begin{abstract}
We investigate the details of the electronic structure in the neighborhoods
of a carbon atom vacancy in graphene by employing magnetization-constrained
density-functional theory on periodic slabs, and spin-exact, multi-reference,
second-order perturbation theory on a finite cluster. The picture
that emerges is that of two local magnetic moments (one $\pi$-like
and one $\sigma$-like) decoupled from the $\pi$ band and coupled
to each other. We find that the ground state is a triplet with a planar
equilibrium geometry where an apical C atom opposes a pentagonal ring.
This state lies $\sim$0.2 eV lower in energy than the open-shell
singlet with one spin flipped, which is a bistable system with two
equivalent equilibrium lattice configurations (for the apical C atom
above or below the lattice plane) and a barrier $\sim$0.1 eV high
separating them. Accordingly, a bare carbon-atom vacancy is predicted
to be a spin-one paramagnetic species, but spin-half paramagnetism
can be accommodated if binding to foreign species, ripples, coupling
to a substrate, or doping are taken into account.
\end{abstract}
\maketitle

\section{Introduction}

Magnetism in graphene is a fascinating and highly controversial matter\cite{katsnelson-book}.
Early reports on ferromagnetic ordering in graphite and graphene\cite{Esquinazi02,Esquinazi03,Esquinazi07,wang09}
have been questioned in the light of the ubiquitous presence of magnetic
contaminants, and measurements under carefully controlled conditions
showed that graphene, like graphite, is strongly diamagnetic with
a weak paramagnetic contribution from adatoms and/or carbon atom vacancies\cite{Sepioni10}.
Simple adsorbates such as fluorine and missing carbon atoms have been
shown to provide a spin-$1/2$ paramagnetic response\cite{Nair12},
though spin-$1$ paramagnetism has been reported upon N$^{+}$ irradiation\cite{ney11}.
Direct evidences of the presence of adatom- and vacancy-related magnetic
moments have been observed in pure spin transport measurements on
graphene spin valves\cite{McCreary12}. Likewise, signatures of Kondo
effect have been observed in charge-transport measurements at low
temperatures on irradiated graphene, and high Kondo temperatures reported
for finite and zero electron density\cite{fuhrer11}, at odds with
the above magnetometry measurements\cite{Nair12}.

From a theoretical perspective, perfect bipartite systems support
a number of zero-energy {}``midgap'' states which is greater or
equal than the sublattice imbalance $|n_{A}-n_{B}|$, where $n_{A}$,$n_{B}$
are the number of sites in the two sublattices\cite{Inui1994,Fajtlowics}.
When imbalance results from isolated missing $p_{z}$ orbitals (\emph{e.g.}
for low concentrations of covalently bound adatoms or vacancies) these states decay slowly
($\sim1/r$) from the defects and localize on the locally majority sites\cite{Pereira2006,pereira08a}, 
as confirmed by experiments\cite{Ugeda10}. Thus, these defects form quasi-localized
$\pi$ moments, which couple either ferromagnetically or antiferromagnetically
depending on their lattice position. In fact, with local interactions
only, at charge neutrality (half-filling) the spin state of the system
exactly matches the sublattice imbalance\cite{TheoLieb}, $S=|n_{A}-n_{B}|/2$,
and thus coupling is ferromagnetic for defects in the same sublattice
and antiferromagnetic otherwise, a result which also follows by analyzing
RRKY interactions in graphene\cite{Kogan11}. Within the same assumptions
(perfect electron-hole symmetry, local interactions only) coupling
between $\pi$ moments and conduction states has been investigated
beyond mean-field approaches and found to be ferromagnetic\cite{Guinea11},
thereby confirming that simple adatoms covalently bound to the substrate
(\emph{e.g.} H, F species) behave as spin-$1/2$ localized moments.
In turn, this also affects chemical properties and favors formation
of dimers of balanced type\cite{Hornekaer2006a,casolo09}. 

This simple picture has to be revised for a carbon atom vacancy where,
in addition to the above $\pi$ midgap state, three $\sigma$ orbitals
are left singly occupied upon vacancy formation, and a structural
instability (Jahn-Teller distortion) arises which breaks electron-hole
symmetry, even if nearest neighbors interactions only are retained.
The ensuing lattice re-arrangement leaves two unpaired electrons,
and a magnetic moment in the range $2.0-1.0\,\mu_{B}$ has been found
by (ensemble) density functional theory (DFT) calculations\cite{Elbarbary03,Lethinen2004,Yazyev07,Dharmawardana200880,Dai11,palacios12},
with a clear tendency to $1.0\,\mu_{B}$ (and vanishing dependence
of the energy on the magnetization) in the low-density limit which
signals the absence of any order at experimentally relevant concentrations\cite{palacios12}.
Apart from the possible role of electron-hole symmetry, this apparently
contrasts with the situation described above for a $\pi$ moment only,
since the presence of an additional unpaired electron in a very localized
$\sigma$ orbital is expected to give either a triplet or a singlet
state. Recent experiments have shown that the spin$-1/2$ paramagnetism
of missing carbon atoms has two contributions\cite{Nair13}, from
$\sigma$ and $\pi$ states respectively, and one of them can be quenched
upon molecular doping and possibly by means of the electric field
effect\cite{Nair13}. This result requires that, at least under the
experimental conditions of Ref. \onlinecite{Nair13}, the unpaired
electrons around a vacancy negligibly interact with each other, in
contrast with early reports on spin-1 paramagnetism of irradiated
graphene samples\cite{ney11}. Proper consideration of $\sigma$ states,
and their possible hybridization with $\pi$ states when the substrate
is no longer locally planar, \emph{e.g.} because of ripples or interaction
with a substrate\cite{ugeda11}, has led to reconsidering the issue
of the interaction between the localized magnetic moments and the
conduction electrons\cite{Guinea12,Fritz12b}, though the most recent
experiments\cite{Nair13} seem to rule out this possibility.

Here, in order to help shed light on the above issues we consider
in detail the electronic structure around a carbon atom vacancy in
graphene. We employ both conventional DFT methods in periodic models
and accurate, spin-exact quantum chemistry methods in a finite cluster,
to investigate the coupling of the two electrons left unpaired after
vacancy formation and reconstruction. We analyze several substrate
geometries close to the equilibrium one, and focus in particular on
the out-of-plane movement of the carbon atom where most of the unpaired
electron density resides. Our results show that the triplet is the
ground-state and has a planar equilibrium geometry, while the singlet
--which lies $\sim0.2$ eV above it-- gets easily stabilized by an
out-of-plane lattice distortion. Hence, both spin-1 and spin-$1/2$
paramegnetism may in principle arise in irradiated graphene, depending
on local interactions, curvature, etc. of the graphene sheet, in addition
to doping or chemical interactions with foreign species.

The manuscript is organized as follows: Section \ref{sec:Jahn-Teller-distortion}
outlines the important Jahn-Teller distortion occurring in the system
and Section \ref{sec:methods-and-models} gives the details of the
electronic structure methods adopted in this work. Results are given
in Section \ref{sec:Results-and-discussion} and discussed in Section
V, and Section VI summarizes and concludes.

\section{Jahn-Teller distortion\label{sec:Jahn-Teller-distortion}}

\begin{figure}
\begin{centering}
\includegraphics[width=0.8\columnwidth]{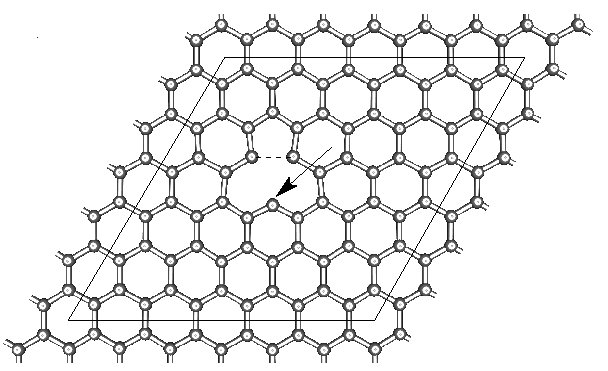}
\par\end{centering}

\caption{\label{fig:Optimized-structure-of}Optimized structure of a carbon
vacancy in a 6x6 unit cell, with the apical carbon marked by an arrow. }

\end{figure}

As mentioned above, formation of a carbon atom vacancy gives rise
to localized states around the vacancy, namely one $\pi$ (semilocalized)
{}``midgap'' state and three dangling orbitals in the $\sigma$
network which result from breaking the $sp^{2}$ bonds which hold
the carbon atom in place. In the local $D_{3h}$ point symmetry group
which is appropriate to discuss proper and pseudo Jahn-Teller distortions,
the first belongs to $a_{2}^{"}$ symmetry species, and the latter
span $a'_{1}+e'$ irreducible representations, $a'_{1}$ being lowest
in energy since it contains a purely bonding combination of $\sigma$
orbitals. Hence, the lowest energy {}``scenarios'' for the many-body
electronic state can be obtained by distributing two electrons in
the $e'$ and $a_{2}''$ states, \emph{i.e.} starting from configurations
of the type $..(a'_{1})^{2}(e')^{n_{1}}(a{}_{2}'')^{n_{2}}$ with
$n_{1}+n_{2}=2$. Among these, the one with $n_{1}=n_{2}=1$ is expected
to be lowest in energy and gives rise to many-body states of $E''$
symmetry for both the parallel and antiparallel alignment {[}The remaining
possibilities with two electrons in the same set of states are pushed
up in energy by a larger Coloumb repulsion and have symmetries $^{1}A'_{1}+^{3}A'_{2}+^{1}E'$
for $n_{1}=2$ and $^{1}A'_{1}$ for $n_{2}=2${]}. Thus, the ground-state
is doubly degenerate for both spin alignments and undergoes (proper
or pseudo) Jahn-Teller distortion. The latter occurs because of coupling
with in-plane $e'$ vibrations ($[E'']^{2}=[E']^{2}=A'+E'$) which
distort the symmetric arrangement of the carbon atoms around the vacancy.
This is a standard $E\otimes e$ problem which is described by the
so-called \emph{tricorn} when such vibrations are included up to second-order\cite{bersuker01,bersuker-book}.
Accordingly, a distorted geometry with a {}``pentagonal'' ring and
a {}``apical'' carbon atom opposite to it (see Fig.\ref{fig:Optimized-structure-of}),
can be predicted to be the equilibrium configuration (three-fold degenerate),
as indeed found in several previous investigations\cite{Elbarbary03,Lethinen2004,Yazyev07,Dharmawardana200880,Dai11,palacios12}. 

Out of plane, $E''$ vibrations do not lift degeneracy at first-order,
but may affect energetics at higher orders, especially if coupling
to the low-lying excited states is considered\cite{bersuker-book}.
This is particularly important here since such distortions are qualitatively
different for the states $..\pi^{1}\sigma^{1}$ depending on whether
the spins are parallel or antiparallel to each other. This is evident
for the out-of-plane movement of the apical carbon atom in the distorted
configuration: the $\sigma$ and the $\pi$ states may hybridize to
some extent (and gain energy from double filling) if the two electrons
couple at low-spin, otherwise they require extra energy to reduce
their overlap. As a consequence, the planar structure is expected
to be stable in the triplet state only; in the singlet, a non-planar
configuration with the apical carbon atom slightly above (or below)
the surface plane appears to be more stable. Because of that, the
relative stability of the two spin states is geometry-dependent and
its analysis requires at least investigating the out-of-plane movement
of the {}``apical'' carbon atom. This is described in Section \ref{sec:Results-and-discussion},
after Section \ref{sec:methods-and-models} has introduced the electronic
structures methods and setups adopted, along with the structural models
chosen to investigate the vacancy.

\section{methods and models\label{sec:methods-and-models}}

Electronic structure calculations were performed at different correlation
levels for different structural models. Periodic arrangements of vacancies
in large unit cells were investigated with standard, plane-wave based
density functional theory calculations, whereas a finite-size (cluster)
model was judiciously selected and studied with correlated wavefunction
methods described below.

\subsection{Periodic models }

\begin{table}
\begin{centering}
\begin{tabular}{|c|c|c|c|c|c|c|}
\hline 
$n$ & $\triangle E$ ($meV$) &  $M_{p}(\mu_{B})$  & $d_{CC}^{p}($\AA$)$ & $M_{np}(\mu_{B})$ & $d_{CC}^{n}($\AA$)$ & $h_{C}($\AA$)$\tabularnewline
\hline
\hline 
$4$ & 27.0 & 1.642 & 2.212 & 0.411 & 2.232 & 0.25\tabularnewline
\hline 
$5$ & 38.2 & 1.889 & 2.126 & 0.111 & 2.169 & 0.28\tabularnewline
\hline 
$6$ & 36.3 & 1.556 & 2.007 & 0.444 & 2.075 & 0.25\tabularnewline
\hline 
$7$ & 30.2 & 1.556 & 1.999 & 0.444 & 2.026 & 0.24\tabularnewline
\hline 
$8$ & 28.3 & 1.556 & 1.985 & 0.450 & 1.958 & 0.23\tabularnewline
\hline 
$9$ & 26.9 & 1.556 & 1.978 & 0.446 & 1.969 & 0.23\tabularnewline
\hline 
$10$ & 27.9 & 1.556 & 1.962 & 0.463 & 1.952 & 0.25\tabularnewline
\hline
\end{tabular}
\par\end{centering}

\caption{\label{tab:Full magn}Results of full structural relaxation without
constraints on the magnetization, for a vacancy in several $n\times n$
supercells. $\Delta E$ is the energy separation between the metastable
non-planar ($C_{s}$) structure and the planar ($C_{2v}$) minimum,
$M_{np}$ is the magnetization of the former and $M_{p}$ that of
the latter. Also reported the length of newly formed $CC$ bond closing
the pentagon ($d_{CC}^{p}$ and $d_{CC}^{n}$ for planar and non-planar
geometries, respectively) and the height $h_{C}$ of the apical carbon
atom in the non-planar configuration. }
\end{table}
Periodic models were studied with plane-wave DFT as implemented in
the Vienna \emph{ab initio} package suite (VASP)\cite{VASP1,VASP4}.
The exchange-correlation effects were treated with the Perdew-Burke-Ernzerhof
(PBE)\cite{PBE1,PBE2} functional within the generalized gradient
approximation (GGA), in the spin-polarized framework. Kohn-Sham orbitals
were expended on a plane-wave set limited to a $500$ eV energy cutoff
and core electrons were frozen and replaced by projector-augmented
wave (PAW) pseudopotentials\cite{PAW1,PAW2}. Several $n\times n$
graphene supercells with a $20\:\textrm{\AA}$ vacuum were initially
considered to model the defective system, from $n=2$ to $n=10$,
by using $\Gamma$ centered $k-$point meshes ranging from $15\times15\times1$
(for $n=2$) to $3\times3\times1$ for $n=6-10$. The structure of
a vacancy in such cells was fully optimized without constraints on
the magnetization and gave a Jahn-Teller distorted planar minimum,
with a $C-C$ bond length in the pentagon of about $2$ \AA, \emph{i.e.}
smaller than the graphene lattice constant $a=2.46\:\textrm{\AA}$  
but much larger than a typical (single) $C-C$ bond ($1.54\:\textrm{\AA}$).
Additionally, starting with a low-magnetization guess, we invariably
found for $n\ge4$ a \emph{metastable} spin-polarized solution with
vanishing magnetization and a non-planar configuration. Its energy
separation $\Delta E$ from the planar minimum in the same supercell,
along with the resulting magnetizations $M$ and the main geometrical
parameters of the two structures are given in Table \ref{tab:Full magn}.
Results for the minimum are in good agreement with those found in
previous studies\cite{Elbarbary03,Lethinen2004,Yazyev07,Dharmawardana200880,Dai11,palacios12}. 

On the basis of these results we concluded that a $6\times6$ supercell
with a $6\times6\times1$ $k-$point mesh was a good compromise between
the need of reducing interaction between periodic images and computational
manageability. Therefore, further investigations were performed with
this setup.

\subsection{A finite-size model}

\begin{figure}
\begin{centering}
\includegraphics[width=0.4\columnwidth]{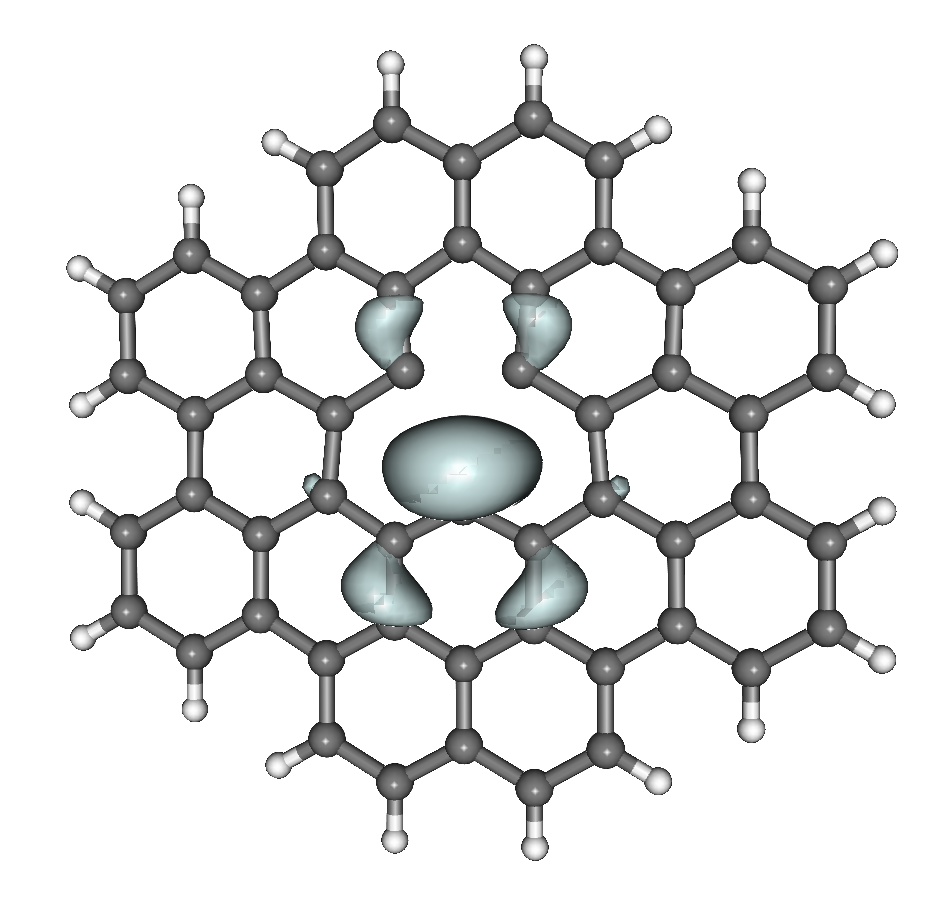}\includegraphics[width=0.4\columnwidth]{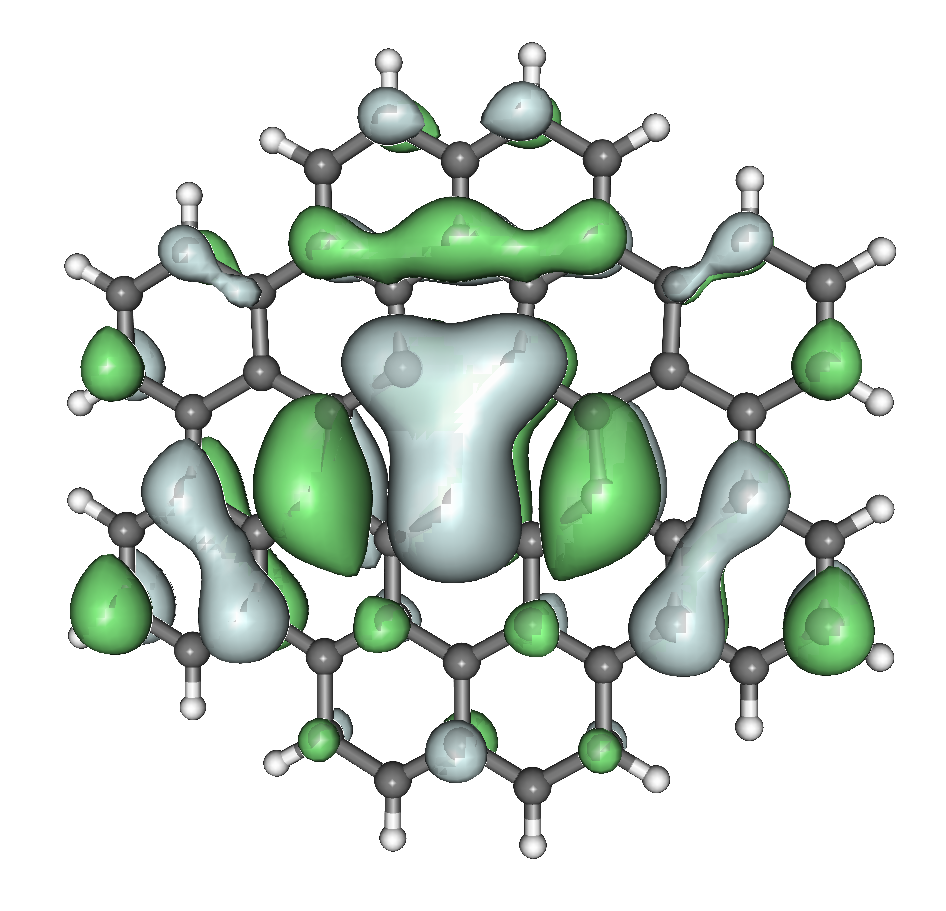}
\par\end{centering}

\caption{\label{fig:The-molecular-model}The molecular model adopted for the
wavefunction calculations, along with isosurfaces of singly occupied
$\sigma$ (left) and $\pi$ -midgap (right) orbitals for $|\phi|=0.015$
\AA$^{-3/2}$, as obtained at the restricted open-shell Hartree-Fock
level ($S=1$) for the minimum structure. }
\end{figure}
 The semi-localized character of the electronic states induced by
the vacancy makes it realistic the study of finite-size models where
accurate, correlated wavefunction calculations are feasible. The size
(and shape) of the cluster has to be large enough to minimize the
effects that edges have on the details of the electronic structure,
and small enough that complex many-body wavefunctions are yet tractable.
To this end we considered a reasonably sized Polycyclic Aromatic Hydrocarbon
(PAH) ($C_{53}H_{20}$), a carbon cluster with a central vacant site
which is hydrogen-terminated at the edges (Fig. \ref{fig:The-molecular-model}).
Its actual shape was chosen -following the line of reasoning of Ref.
\onlinecite{bonfanti11}- with the help of Tight-Binding (TB) calculations
in such a way to limit the edge localization which does interfere
with the defect-induced states at the Fermi level. In the chosen structure,
edge states were found sufficiently far in energy from the vacancy-induced
states (both at the TB and at the Hartree-Fock (HF) level) to make
us confident that the resulting energetics accurately describes the
vacancy. Cluster geometries were selected with a `cut-out process'
starting from the above mentioned $6\times6$ supercells, and adding
hydrogen atoms to the undercoordinated edge C atoms, without further
geometry refinement. In this way, comparison between the periodic
and the cluster model with the same local arrangement close to the
vacancy was possible. 

Accurate results on the finite model were achieved through all-electron,
correlated wavefunction calculations based on atom-centered basis-sets
of the correlation consistent type (cc-pVDZ). Energy was obtained
with the help of the MOLPRO suite of codes\cite{MOLPRO_brief} by correcting
to second order in perturbation theory a `reference' wavefunction
of the Complete Active Space Self-Consistent Field (CASSCF) type,
according to what is known as CASPT2\cite{molpro-caspt2-1,molpro-caspt2-2}.
The CASSCF($n$,$m$) wavefunction is a multi-determinant wavefunction
containing all possible excitations of $n$ `active' electrons in
$m$ `active' orbitals, where all the orbitals and expansion coefficients
are variationally optimized\cite{MOLPRO-WK85,MOLPRO-KW85}. For our
purposes, we started with a minimal active space containing the $\sigma$
and $\pi$ orbitals localized on the vacancy and the two electrons
occupying them at the HF level, and enlarged it by including two further
$\pi$ orbitals (one below and one above the Fermi level), \emph{i.e.}
CAS(4,4). We fully optimized the active orbitals and the thirty doubly-occupied
orbitals higher in energy, and kept the lowest-lying (doubly occupied)
orbitals frozen at the Hartree-Fock level. With the optimized CASSCF
wavefunctions at hand, dynamic correlation was introduced by including
perturbatively the effect of single and double excitations out of
the configurations contained in the selected CAS `reference' space.

\section{Results\label{sec:Results-and-discussion}}

\subsection{Periodic calculations}

As shown in Tab.\ref{tab:Full magn} all the optimized structures
show an appreciable magnetization, in agreement with previous results
obtained for similar systems\cite{Elbarbary03,Lethinen2004,Yazyev07,Dharmawardana200880,Dai11,palacios12},
and this is significantly different for the two structures. The two
geometries thus refer to the different electronic states -the high
and the low spin state, respectively- which originate from the different
ways the unpaired electrons localized on the vacancy align. Unfortunately,
even assuming that the KS wavefunction correctly describes the spin
state of the system, proper assignment is not possible with DFT. This
is because (i) allowance of fractional occupation of single-particle
Kohn-Sham states, as occurs with ensemble-DFT, typically prevents
the appearance of pure states even when they are in principle possible
at $T=0$ K, and (ii) pure but spin-polarized (unrestricted) solutions
do \emph{not} have a definite spin value ($S^{2}$).

In order to get rid of these problems, and to mimic as much as possible
the expected electron configurations, we relied on magnetization-constrained
DFT calculations on the $6\times6$ supercell, setting the (projection
of the) magnetic moment to two (zero) Bohr magnetons for the triplet
(singlet) case. Full structural optimizations were then performed
for different out-of-plane displacements $h$ of the apical carbon,
for each `spin' state, to investigate how these states evolve out
of the plane.%
\begin{figure}
\begin{centering}
\includegraphics[clip,width=0.65\columnwidth]{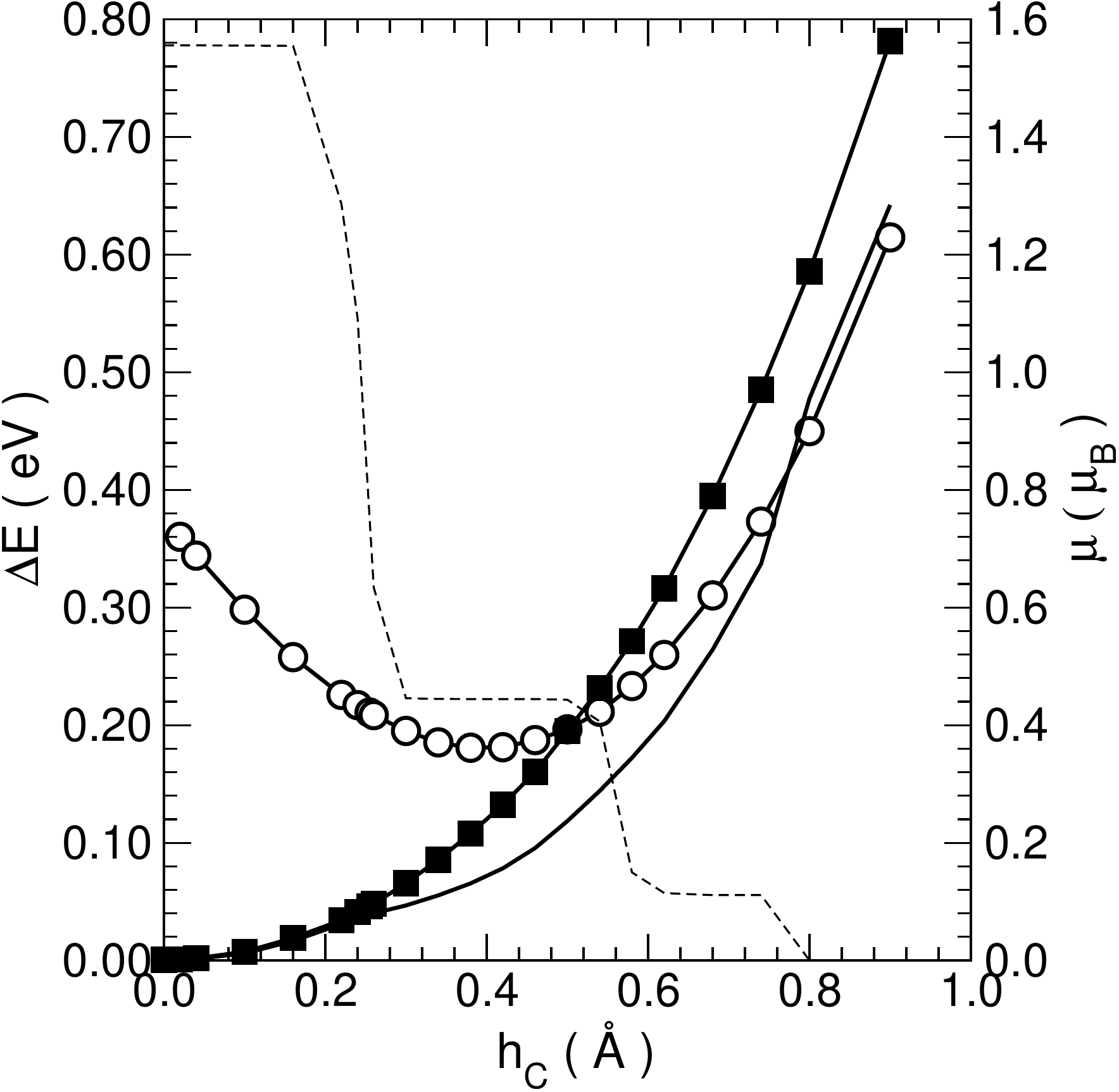}
\par\end{centering}

\caption{\label{fig:M energies}Magnetization-constrained energies as functions
of the height $h$ of the apical carbon atom above the surface. Filled
and empty symbols for $M=2,0\,\mu_{B}$, referenced to the minimum
of the $M=2\,\mu_{B}$ case. Also reported as a thick line the results
of magnetization-unconstrained calculations, referenced to their minimum,
and the corresponding magnetization (dashed line, right scale). }
\end{figure}

The results of such calculations are shown in Fig. \ref{fig:M energies},
referenced to the planar configuration in the triplet state, along
with the magnetization-unconstrained curve referenced to its minimum
(which is only $29$ meV below that of the constrained triplet curve).
It is clear that the latter is a mixture of the two electronic states,
with the triplet prevailing for $h\approx0$ and the singlet dominant
for $h\gg0$. 

From Fig. \ref{fig:M energies}, we also see that the global minimum
belongs to the triplet curve and has a flat geometry (of symmetry
$C_{2v}$). The singlet curve instead shows two equivalent minima
(of symmetry $C_{s}$) for the carbon atom above and below the plane,
respectively, $\sim0.4\:\textrm{\AA}$ away from the surface. The
latter thus represents a bi-stable system that crosses the triplet
when the carbon atom moves out by about $\sim0.5\:\textrm{\AA}$,
but is otherwise higher in energy. The energy difference between the
singlet and triplet minima is $\sim0.18$ eV, thus significantly larger
than the (unconstrained) $\triangle E$ for the same $6\times6$ supercell
reported in Table \ref{tab:Full magn}, which referred to `mixed'
electronic states. The singlet minima are separated by a barrier $\sim0.2$
eV high which lies $\sim0.4$ eV above the triplet. This estimate
will be refined in the next section on the basis of more accurate
electronic structure calculations. 
Before leaving this section we
only stress that the curves reported in Fig. \ref{fig:M energies}
refer to a full structural relaxation (in the given electronic state)
with respect to all the degrees of freedom but the height of the apical
carbon atom, and thus different geometries for the triplet and for
the singlet typically result for the \emph{same} $h$ value. The differences
though are minimal as the height of the apical carbon atom is the
main geometrical parameter controlling spin alignment in this system,
hence graphs such as those of Fig. \ref{fig:M energies} are also
representative of vertical energy differences. For instance, the pentagon
CC bond length is $2.035\:\textrm{\AA}$ in the triplet equilibrium
configuration and increases to $2.081\:\textrm{\AA}$ in the singlet
minima, to be compared with $d_{CC}=2.007\:\textrm{ \AA}$ for the
magnetization-free planar structure (Table \ref{tab:Full magn}) and
$d_{CC}=2.467\:\textrm{ \AA}$ in pristine graphene.

\subsection{Wavefunction calculations }

\begin{figure}
\centering{}\includegraphics[clip,width=0.65\columnwidth]{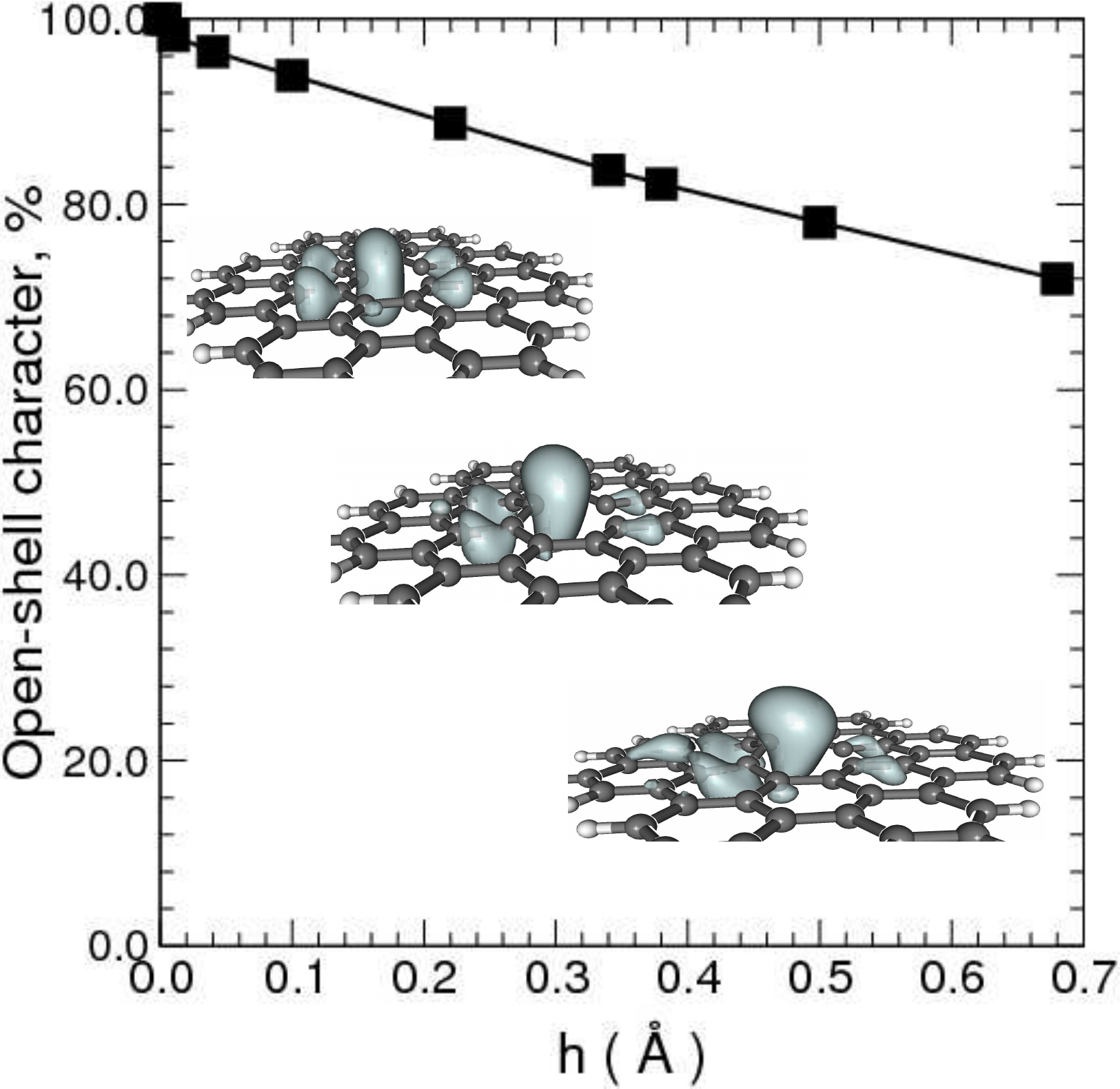}\caption{\label{fig:open-shell}Variation of the open-shell character of the singlet state as the
apical carbon atom moves out of plane. The symbols give the weight
$(c_{1}^{2}+c_{2}^{2})\times100$ of the coefficients in the minimal
CAS(2,2) wavefunction with diabatized orbitals described in the main
text. The insets show isosurfaces at $|\phi|=0.015$ \AA$^{-3/2}$
of the corresponding $\sigma$-like CASSCF orbitals for representative
values of $h=0.0,\,0.32,\,0.68$ \AA, from left to right, respectively. }
\end{figure}
To analyze in detail the electronic structure in the neighborhoods
of the vacancy we performed additional wavefunction calculations,
as described in Section \ref{sec:methods-and-models}, on the geometries
obtained at the DFT level, fit to the cluster model of Fig. \ref{fig:The-molecular-model}.
The use of a cluster introduces a number of issues related to the
finiteness of the model (\emph{e.g.} dependence on the system size,
discretization of the continuum of states, etc. ) but removes some
oddities of the supercell approach, \emph{e.g.} the periodic repetition
of defects in the same sublattice which would favor ferromagnetic
alignment. We first checked that the singlet had the correct open-shell
character at the planar geometry, namely that the wavefunction reads
approximately as \[
\Psi_{h=0}^{S}\propto|...\phi_{\sigma}^{\alpha}\phi_{\pi}^{\beta}|\pm|...\phi_{\sigma}^{\beta}\phi_{\pi}^{\alpha}|\]
where $\phi_{\sigma}$ and $\phi_{\pi}$ are $\sigma$-like and $\pi$-like
orbitals on the apical carbon atoms, respectively, $|..|$ is a shorthand
for a Slater determinant and the plus (minus) sign holds for the triplet
(singlet) state. For non-planar geometries $\sigma-$like and $\pi-$like
orbitals get generally mixed, and the singlet displays both open-
and closed-shell character. Only if the first dominates the singlet
can be considered to be the {}``same'' electronic state of the triplet
but with one spin flipped, and the singlet-triplet energy separation
is meaningful of an exchange coupling. 

To check this, we exploited the invariance of the CASSCF wavefunction
with respect to rotations of the active orbitals, and choose orbitals
which maximize the overlap (while keeping orthogonality) with the
above $\phi_{\sigma}$ and $\phi_{\pi}$ states of the planar case.
In this case \[
\Psi_{h}^{S=0}\approx c_{1}|...\phi_{\sigma}^{\alpha}\phi_{\pi}^{\beta}|+c_{2}|...\phi_{\sigma}^{\beta}\phi_{\pi}^{\alpha}|+c_{3}|...\phi_{\sigma}^{\alpha}\phi_{\sigma}^{\beta}|+c_{4}|...\phi_{\pi}^{\alpha}\phi_{\pi}^{\beta}|\]
where $c_{1}$ and $c_{2}=-c_{1}$ represent the open-shell contribution,
and $c_{3}$ and $c_{4}$ account for the closed-shell character ($c_{3}=c_{4}=0$
and $c_{2}=c_{1}$ in the triplet). In Fig. \ref{fig:open-shell}
we report the evolution of the weight $|c_{1}|^{2}+|c_{2}|^{2}$ in
the normalized wavefunction, as a measure of the open-shell character
in singlet state. They were obtained from simple CAS(2,2) calculations
on the singlet-optimized geometries -\emph{i.e.} using just the four-determinant
wavefunction described above- but similar results are found for more
elaborate functions. Evidently, the system is of open-shell type for
a wide range of $h$ values, comprising the equilibrium one. Only
for very large values of $h$ the system prefers a closed-shell configuration
with {}``magnetic'' properties turned off; correspondingly the triplet
is pushed much higher in energy. 

\begin{figure}
\centering{}\includegraphics[clip,width=0.6\columnwidth]{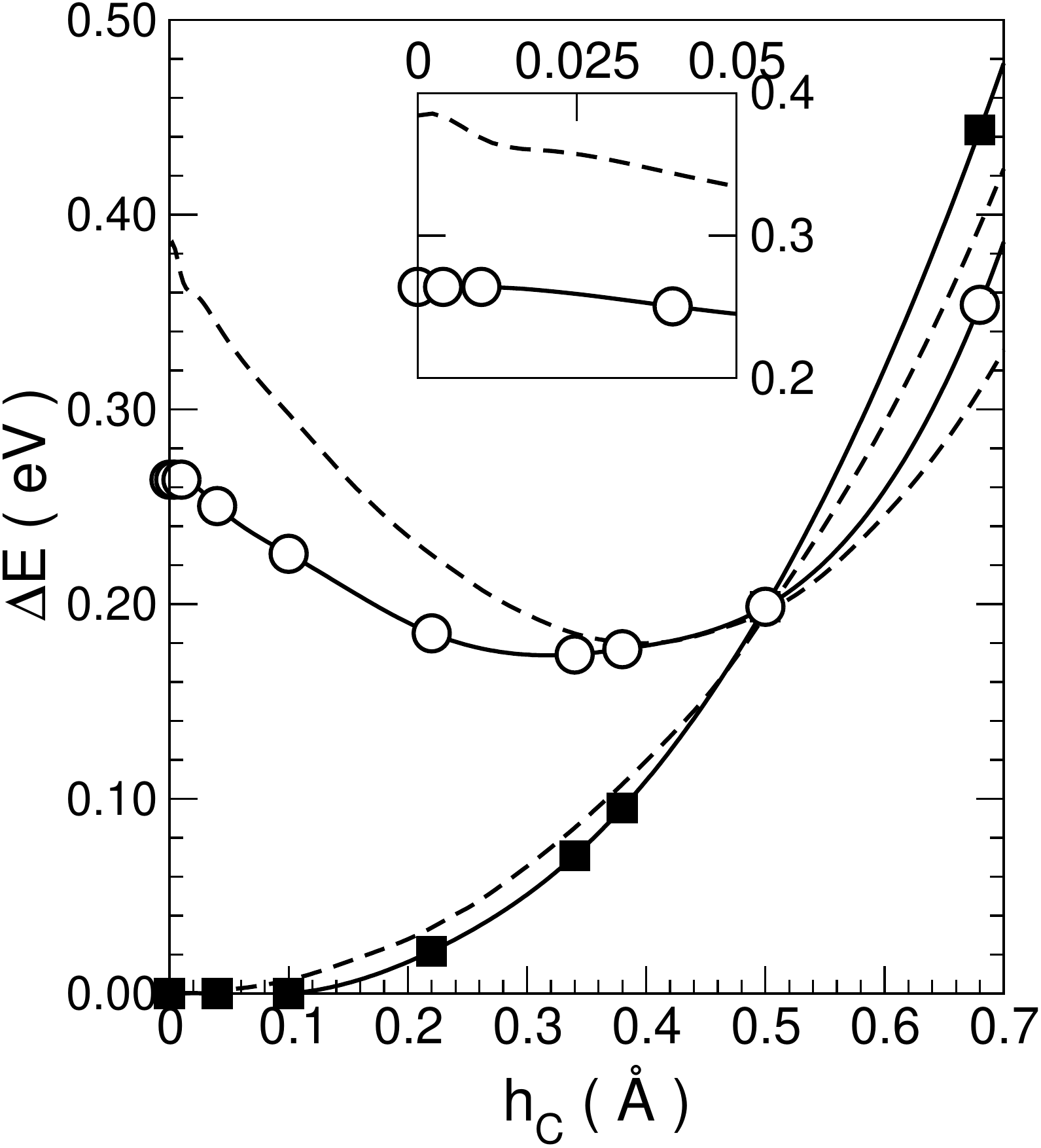}\caption{\label{fig:CASPT2-energy}CASPT2 energies as function of the displacement
($h$) of the apical carbon atom out of the surface plane, for the
singlet (open symbols) and the triplet (filled symbols) states. Also
reported for comparison the magnetization constrained DFT results
of Fig.\ref{fig:M energies} (dashed lines). Energies are referenced
to the triplet minimum at the corresponding theory level. The inset
shows a blow-up of the $h\approx0$ region for the singlet.}
\end{figure}

Multi-configuration SCF wavefunctions obtained distributing 4-electrons
in 4-orbitals (CAS(4,4)) were then optimized for several geometries
sampled from the singlet and the triplet curves in Fig. \ref{fig:M energies},
and used as references for perturbative (CASPT2) calculations. The
results are shown in Fig. \ref{fig:CASPT2-energy} , together with
DFT ones for comparison, for several values of the $h$ coordinate,
referenced to the triplet minimum.%

It is evident from Fig. \ref{fig:CASPT2-energy} that CASPT2 and DFT
results closely parallel each other for the triplet but differ substantially
in the singlet. In the latter case, a cusp (due to a likely interaction
with higher lying electronic states) is only present at the DFT level,
and smooth out at the CASPT2 level, thereby signalling the presence
of an avoided crossing. In fact, the CAS(4,4) space is sufficiently
large to allow us to properly describe a number of quasi-degenerate
states, \emph{i.e.} those obtained by placing all the four unpaired
electrons of the vacancy in low-lying states. 

As expected, ferromagnetic coupling is preferred for most values of
the height of the carbon above the surface, and a crossing results
at about $h=0.5\:\textrm{\AA}$; for larger values of $h$, the gain
in hybridization energy overcomes Coulomb repulsion, and the system
show increased close-shell character. A minimum occurs in the singlet
at $h=h_{0}=0.38\:\textrm{\AA}$ ($h_{0}\approx0.4\:\textrm{\AA}$
at the DFT level), and is actually a double minimum, $h=\pm h_{0}$,
on account of the meaning of the $h$ coordinate. The singlet is thus
a symmetric bistable system with a barrier height of $E_{b}=0.09$
eV. \emph{}

\begin{figure}
\centering{}\includegraphics[clip,width=0.5\columnwidth]{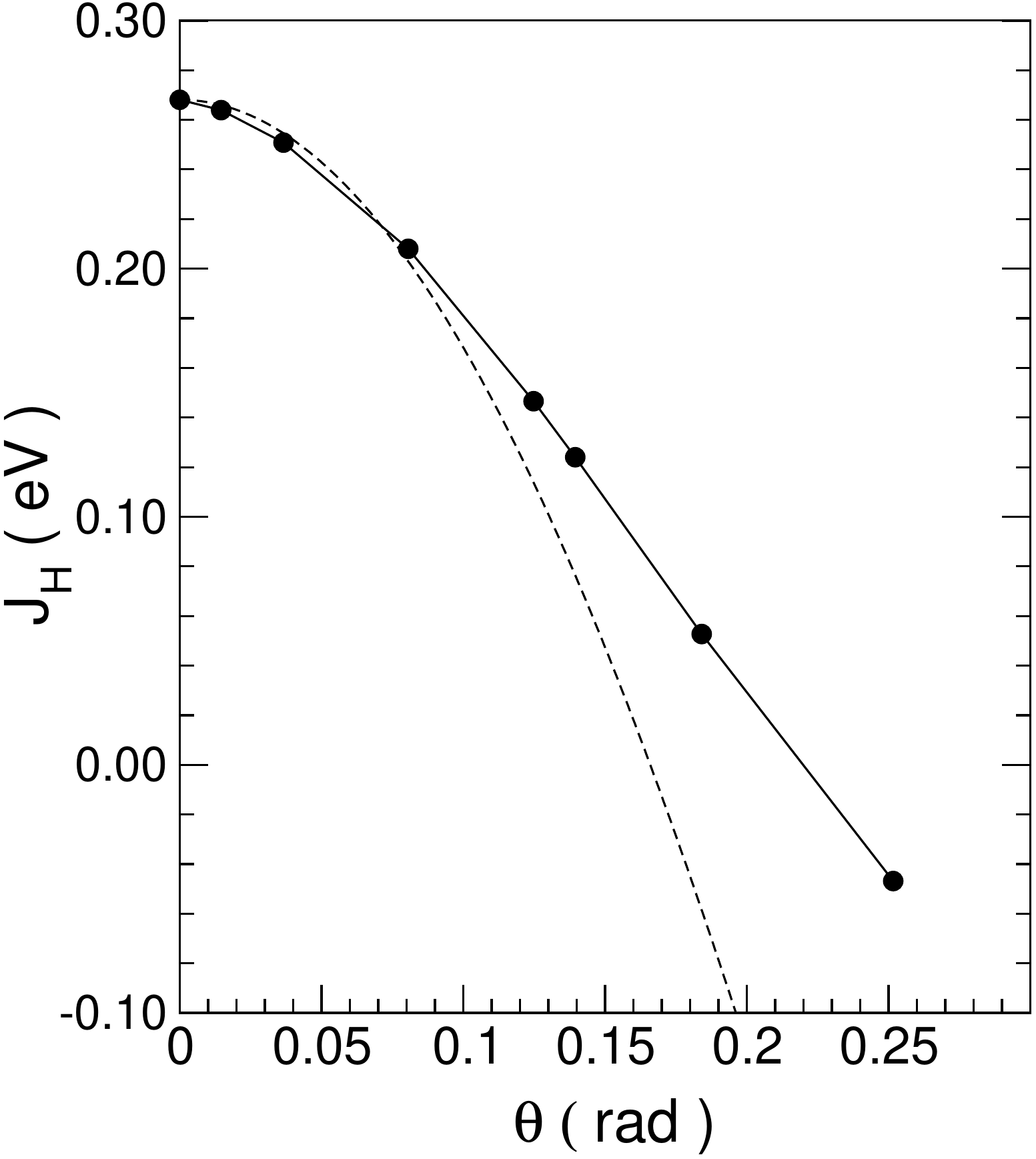}\caption{\label{fig:Hund}Variation of the Hund coupling constant as a function
of the angle subtended by the $\sigma$ dangling bond and the graphene
plane (filled symbols). Dashed line is a low-angle weighted-fit of
the data to $J_{H}=J_{H}^{0}-4A\, tg^{2}(\theta)\left(1-2tg^{2}(\theta)\right)/3$
, which gives $J_{H}^{0}=0.268$ eV and $A=7.59$ eV.}
\end{figure}

Exchange (Hund) coupling was obtained by the vertical singlet-triplet
energy separation, $J_{H}=\Delta E_{ST}$, using the geometries optimized
for the triplet, and is shown in Fig.\ref{fig:Hund} as a function
of the angle $\theta$ subtended by the $\sigma$ dangling bond and
the graphene plane. The results closely parallel those reported in
Fig. \ref{fig:CASPT2-energy} since, as observed above, the difference
between singlet and triplet geometries are minimal. The coupling ranges
from $J_{H}\sim0.25$ eV in the planar configuration to about $J_{H}\sim0.1$
eV in the equilibrium configuration of the singlet, and thus $J_{H}\sim0.25-0.20$
eV seems to be appropriate for the ground-state system. 

It is worth stressing at this point, that $J_{H}$ defined in this
way is the Hund coupling constant related to the geometry-dependent
$\sigma$-like and $\pi$-like orbitals hosting the unpaired electrons.
Its value in the planar structure, $J_{H}^{0}=J_{H}(\theta\equiv0)$,
gives the Hund coupling constant in the Anderson impurity model for
the vacancy\cite{Guinea12}, while its dependence on $\theta$ (at
small angles) simply reflects the behavior of the hybridization strength\cite{pauling-book}
$V_{\sigma\pi}=\sqrt{2}tg(\theta)\sqrt{\left(1-2tg^{2}(\theta)\right)/3}\,\Delta\epsilon_{sp}$
($\Delta\epsilon_{sp}$ being the carbon $s-p$ splitting), as confirmed
by the dashed line in Fig.\ref{fig:Hund}. 

Furthermore, despite the limited size (and discreteness of the energy
spectrum) we found no indication that the $\pi$-midgap state is only
marginally occupied, thereby suggesting that the limit $U_{\sigma\pi}\ll J_{H}/4$
applies in the above mentioned Anderson model ($U_{\sigma\pi}$ is
the Coloumb repulsion between electrons in the $\sigma$ and in the
$\pi$ midgap state). However, there exists the possibility that increasing
the cluster size the situation reverts and the $\pi$-midgap state
depopulates in favour of some other low-lying $\pi$-state ($J=1/2$
problem). %

\section{Discussion\label{sec:Discussion}}

Computed exchange coupling constants are clearly too large to have
a decoupled response from the two localized electrons to external
magnetic fields. The presence of a low-lying singlet at energy $\Delta$
above a $J$-paramagnetic ground-state does affect the magnetization,
but only to the extent it modifies thermal populations, that is introducing
a temperature- and field- dependent correction factor $f(\beta,H)=A(Be^{-\beta\Delta}+A)^{-1}$
to the thermally averaged magnetic moment (here $A=\mbox{sinh}[\beta\gamma H\left(J+\frac{1}{2}\right)]$
and $B=\mbox{sinh}[\frac{\beta\gamma H}{2}]$, $\beta=1/k_{B}T$ as
usual, $H$ is the magnetic field and $\gamma$ is the relevant gyromagnetic
ratio). This factor has a distinguishing feature of making the moment
no longer dependent on the reduced field $\beta H=H/k_{B}T$ only,
but is hardly appreciable for $\beta\Delta\gtrsim1$ (\emph{i.e.}
$T\lesssim2000$ K(!) for $\Delta\sim0.2$ eV). Only for $\beta\Delta\ll1$
this factor transforms the $J=1$ ground-state magnetization density
into twice that of a $J=1/2$ moment. In practice, this limit attains
only if $\Delta$ is vanishing small, since the above (2-electron)
{}``atomic'' picture is challenged at much lower temperatures by
thermal excitations out of/into the $\pi$ midgap state%
\footnote{Notice though that thermal excitations are not relevant to magnetometry
experiments since they are performed at very low temperatures. This
is confirmed by the fact that, under similar conditions, flourinated
graphene gives the expected spin$-\frac{1}{2}$ response of $\pi$
moments\cite{Nair12}. Double occupation of the $\sigma$ state, on
the other hand, is prevented by the large Coulomb repulsion\cite{Guinea12},
$U_{\sigma\sigma}\approx10\,$ eV, and the binding energy in this state is substantial\cite{palacios12}, $\epsilon_\sigma\approx-0.75\,$ eV.%
}. Likewise for doping which can be used to tune the $\pi$ state population,
as it has been recently shown by molecular adsorption\cite{Nair13}.

All this suggests that the bare vacancy in free-standing graphene
at low temperature should display a $J=1$ paramagnetic response and
results reported by Ref. \onlinecite{ney11} are consistent with this
picture. Decoupled $\sigma$ and $\pi$ moments, as those observed
under better-controlled conditions by Nair\emph{ et al.}\cite{Nair12,Nair13},
though, are still plausible since the apical carbon atom may be easily
stabilized out of the surface plane (at about $h\sim0.3\:\textrm{\AA}$ where
$J_{H}\approx0$), for instance in the presence of a (weakly-binding)
substrate\cite{ugeda11} or ripples. 

With the same token, spin-half residual moments may arise because
of interaction with foreign species. Vacancies are highly reactive
species which easily saturate their dangling $\sigma$ bond and leave
a $\pi$ magnetic moment only. We have checked this by considering
adsorption of a single hydrogen atom, and found that such a process
is both thermodynamically and kinetically favored at any temperature:
no barrier is indeed found in the adsorption profile and the binding
energy is about twice that found for adsorption on a bulk site. A
deeper analysis of single and multiple hydrogenation comprising static
and dynamical aspects will be presented shortly.

Screening of the magnetic impurity by $\pi-$band states (Kondo effect)
is a more subtle issue. DFT is not able to handle such highly correlated
situations, and the finite model adopted for the wavefunction calculations,
along with the limited excitations included in the wavefunction, prevent
observation of any pairing between the {}``impurity'' and the $\pi$
band states. In the finite size model, such pairing would be signalled
by the presence of singly excited configurations where $\pi$ states
singlet-couple with the impurity $\pi$-midgap or $\sigma$ state.
Our wavefunction does include a number of excitations out of the occupied
states in the CAS space and, perturbatively, excitations of the ``core''
states at the CASPT2 level, but a detailed analysis of the wavefunction
such that presented above at the CASSCF level is out of question.
In the overall triplet state, Kondo singlet-pairing would be signalled
by an increasing delocalization of the spin-density when enlarging
the cluster size but computational cost becomes prohibitive to check
this {[}for the singlet the spin-density vanishes in any case on account
of the vectorial character of this quantity{]}. Notice though that
dynamical mean-field theory with local interactions showed no evidence
of quenching of the $\pi-$related local magnetic moment\cite{Guinea11},
in accordance with observations of spin-$\frac{1}{2}$ paramagnetism
in fluorinated graphene\cite{Nair12}. In the case of vacancies, screening
of the $\sigma$ moment is expected only for non planar geometries
and, if any, is not compatible with Curie-law behavior observed in
Ref.s \onlinecite{Nair12,Nair13}. Metallic Kondo screening of spin-$\frac{1}{2}$
impurities\textbf{ }has been used to explain transport measurements
in irradiated graphene at different doping levels\cite{fuhrer11},
including charge neutrality, but the interpretation has been questioned\cite{weber12b,fuhrer12},
and the observed logarithmic increase of resistivity at low temperatures
related instead to electron-electron interactions in the disordered
system\cite{altshuler80,weber12a}.

\section{Conclusions}

We reported on a detailed analysis of the electronic and geometric
changes that occur upon vacancy formation in graphene, using both
DFT and a high-level quantum chemistry method (CASPT2) to overcome
known limitations of DFT. The picture that emerges is that two local
magnetic moments coupled to each other to give a triplet ground-state,
in accordance with a report of spin-1 paramagnetic species\cite{ney11}.
Spin-half paramagnetism\cite{Nair12,Nair13}, though, can arise in
many instances. Vacancies are highly reactive and easily saturate
their $\sigma$ dangling bond in the presence of foreign species.
Also, ripples or (weak) coupling to a substrate may stabilize a non-planar
configuration of the apical carbon atom, thereby reducing the effective
Hund coupling constant of the two-electron system and decouple the
corresponding local moments. This is the likely source of spin-half
paramagnetic behavior observed in Ref.s \onlinecite{Nair12,Nair13},
where doping has been shown to effectively halve the magnetization
density. 

We could not deal with the possible pairing of the magnetic moment
with the {}``conduction'' $\pi$-band states, because of the limitation
of DFT on one hand and the use of a finite cluster model (and limited
excitation in the wavefunction) on the other hand. At the above level
of theory we do not have indication, however, of such a coupling.
This is consistent with the absence of anomalies in the measured susceptibility
of Ref.s \onlinecite{Nair12,Nair13}, also at finite densities, and
in that computed (for $\pi$-moments only) with dynamical mean field
theory in the presence of local interactions\cite{Guinea11}, and
suggests that further investigation on the transport data measured
by Ref. \onlinecite{fuhrer11} is required for the Kondo effect in
graphene to be unambiguously identified.

\bibliographystyle{apsrev}


\end{document}